\documentclass[10pt,twocolumn,letterpaper]{article}

\usepackage{./3dv_stuff/3dv}
\usepackage{times}
\usepackage{epsfig}
\usepackage{graphicx}
\graphicspath{{./graphics/}}
\usepackage{amsmath}
\usepackage{amssymb}

\usepackage[pagebackref=true,breaklinks=true,letterpaper=true,colorlinks,bookmarks=false]{hyperref}
\hyperbaseurl{https://}

\usepackage[]{algorithm2e}
\usepackage[font=small,labelfont=bf,textfont=sf]{caption}

\usepackage{epstopdf}
\usepackage{enumitem}

\usepackage{bm}

\usepackage{wrapfig}
\usepackage[percent]{overpic}

\newcommand{\bb}[1]{{\bm{\mathrm{#1}}}}
\newcommand{\Rr}{\mathbb{R}}
\newcommand{\tr}{\text{tr}}

\threedvfinalcopy % *** Uncomment this line for the final submission

\def\threedvPaperID{0032} % *** Enter the 3DV Paper ID here

\ifthreedvfinal\fi

\begin{document}

\title{SpectroMeter: Amortized Sublinear \\ Spectral Approximation of Distance on Graphs}

%% for anonymous conference submission please enter your SUBMISSION ID
%% instead of the author's name (and leave the affiliation blank) !!
\author{Roee Litman\\
Tel-Aviv University\\
%{\href{http://goo.gl/7bxfh3}{\tt tau.ac.il/{\textasciitilde}roeelitm}}
{\small \url{tau.ac.il/~roeelitm}}
% For a paper whose authors are all at the same institution,
% omit the following lines up until the closing ``}''.
% Additional authors and addresses can be added with ``\and'',
% just like the second author.
% To save space, use either the email address or home page, not both
\and
Alex M. Bronstein\\
Technion, Israel Institute of Technology\\
%{\href{https://cs.technion.ac.il/~bron}{\tt cs.technion.ac.il/{\textasciitilde}bron}}
{\small \url{cs.technion.ac.il/~bron}}
}

%\maketitle
\twocolumn[{%
	\renewcommand\twocolumn[1][]{#1}%
	\maketitle
	\begin{center}
		\vskip-15pt		
		\hspace{15pt}
		\begin{overpic}[scale=.095,natwidth=5044,natheight=920]{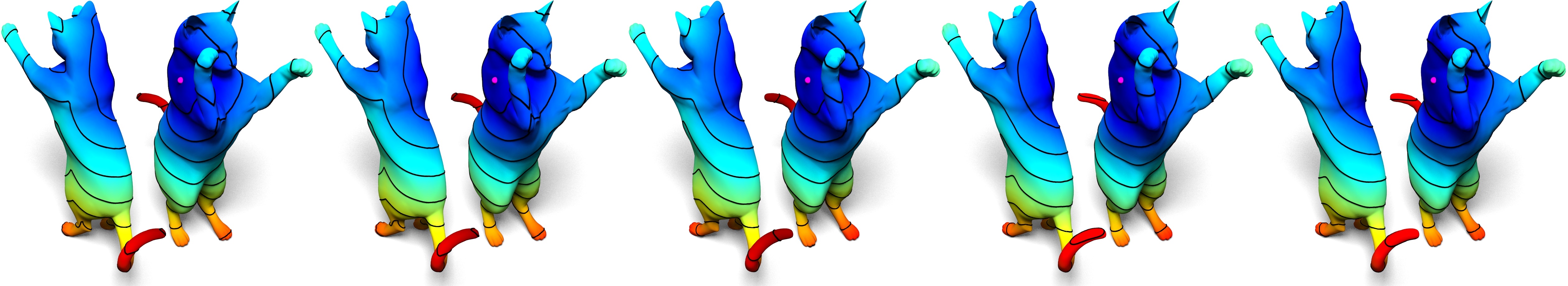}    			
			%\put(00,-5){\parbox{0.12\textwidth}{Pre-calc: \\ Run:}}
			\put(03,-5){\parbox{0.12\textwidth}{\centering Exact\\ polyhedral~\cite{mitchell1987discrete} \\ N/A\footnotemark[1] $\:\diamond\:$ $ 2.3$s }} %\footnote{pre calculation time could not be seperated in the implementation we used}
			\put(22,-5){\parbox{0.12\textwidth}{\centering Fast\\ Marching~\cite{sethian1996fast} \\ $60$ms $\:\diamond\:$ $18$ms }}
			\put(43,-5){\parbox{0.13\textwidth}{\centering Heat\\ method~\cite{Crane} \\ $335$ms $\:\diamond\:$ $7$ms }}
			\put(64,-5){\parbox{0.12\textwidth}{\centering SpectroMeter\\ full \\ $10$s $\:\diamond\:$ $1.5$ms  }}
			\put(84,-5){\parbox{0.13\textwidth}{\centering SpectroMeter\\ sublinear  \\ $15$s $\:\diamond\:$ $0.034$ms }}
			\put(04.5,7){\color{magenta} \circle{2}}
			\put(04.5,7){\color{magenta} \circle{2.2}}
			\put(24.5,7){\color{magenta} \circle{2}}
			\put(24.5,7){\color{magenta} \circle{2.2}}
			\put(44.5,7){\color{magenta} \circle{2}}
			\put(44.5,7){\color{magenta} \circle{2.2}}
			\put(64.5,7){\color{magenta} \circle{2}}
			\put(64.5,7){\color{magenta} \circle{2.2}}
			\put(84.5,7){\color{magenta} \circle{2}}
			\put(84.5,7){\color{magenta} \circle{2.2}}
		\end{overpic}
		\vskip35pt		
		\captionof{figure}{\label{fig:teaser}
			Comparison of five methods for distance map calculations on a triangular mesh from the TOSCA \cite{bronstein2008numerical} dataset.
			Each result is presented from two viewpoints, underwhich the timiming result for pre-calculations (left) and run-time (right) are presented.
			Distances are measured from the magenta point on the cat shoulder, and are colormapped from blue (low) to red (high), and equidistant fronts are marked by black contours.
			The leftmost `exact' map dictates the colormap values on the other four, as well as the locations of the black contours (set at steps of $10\%$ the maximal distance).
			Both SpectroMeter maps (two rightmost) are very similar to the exact ones (two leftmost), but obtained with a fraction of the computational cost.
			Non-differentiable locations are harder to approximate, as in the smoothed cusp is highlighted by a magenta circle.}
	\end{center}%
	%\vspace{2pt}
}]

\begin{abstract}
	We present a method to approximate pairwise distance on a graph, having an amortized sub-linear complexity in its size.
	The proposed method follows the so called heat method due to Crane et al. \cite{Crane}.
	The only additional input are the values of the eigenfunctions of the graph Laplacian at a subset of the vertices.
	Using these values we estimate a random walk from the source points, and normalize the result into a unit gradient function.
	The eigenfunctions are then used to synthesize distance values abiding by these constraints at desired locations.
	We show that this method works in practice on different types of inputs ranging from triangular meshes to general graphs.
	We also demonstrate that the resulting approximate distance is accurate enough to be used as the input to a recent method for intrinsic shape correspondence computation.
	
	\vspace{-10pt}
\end{abstract}

%-------------------------------------------------------------------------
%-------------------------------------------------------------------------
%-------------------------------------------------------------------------

%-------------------------------------------------------------------------
\section{Introduction}

Distances on graphs are used on a daily basis, in many hardware and software applications, sometimes without us being fully aware of it. 
The best known example is map navigation, where one searches the best route to traverse a city, minimizing either distance or commute time.
Another common application is packet routing through the internet, where two of the most commonly used algorithms (OSPF~\cite{moy1997ospf} and IS-IS~\cite{oran1990osi}) allow multiple machines to collaboratively find the best route for a packet.

Two other common domains of application of distance on graphs are computer graphics and geometry processing, where images and shapes are often described as graphs (perhaps, with some additional structure as in the case of meshes).
For example, being able to calculate distances between points on a shape allows to represent the intrinsic information in a way invariant to bending or other extrinsic deformations.

%rendering?
%aligment?
\footnotetext[1]{\label{fnt:pre-calc}Pre-calculation time could not be separated from the runtime in the implementation we used.}

While existing methods for calculating distances on graphs are very efficient, they are still at least linear in the graph size.
This may become prohibitively expensive when the graph is very big or when the number of distance calculations grows.

\vspace{-5pt}
\paragraph{Our contribution.}
We introduce a spectral method to approximate distances on graphs, given only a few of the eigen functions of the graph Laplacian, possibly sampled at a subset of the vertex set.
We see two main contributions of this paper:
First, we show how to perform pairwise distance approximation in sub-linear time in the size of the graph (excluding some preprocessing).
Second, unlike the previous method \cite{Crane}, we do not rely on a discrete \emph{divergence} operator, which allows extension to types of graphs where such an operator is not trivial to define.

%-------------------------------------------------------------------------
%-------------------------------------------------------------------------
%-------------------------------------------------------------------------
\section{Distance maps on graphs}
There exists a wealth of literature dedicated to distance computation on graphs. A useful dichotomy is to categorize such algorithms into \emph{heap based methods} that gradually propagate distances over the graph, usually accessing its vertices in an unpredictable order, and \emph{stack based methods} operating on the entire vertex set or on pre-fixed subsets thereof.

%-------------------------------------------------------------------------
%-------------------------------------------------------------------------
%-------------------------------------------------------------------------
\subsection{Heap based methods}
Heap-based approaches for the computation of distances on graphs have been dominant until, probably, the last decade.
One of the first such algorithms was suggested by Dijksrta~\cite{dijkstra1959note}, solving the shortest path problem for any directed graph with $n$ vertices in $O(n \log n)$.
Mitchell et al.~\cite{mitchell1987discrete} extended this algorithm to polyhedral meshes, providing an exact solution for the discrete geodesic distance computation in $O(n^2 \log n)$. It was later shown by Surazhsky et al.~\cite{surazhsky2005fast} that an approximation with rather low worst-case error bounds can be achieved in $O(n \log n)$.

Since graphs often constitute a discretization of a continuous domain, the question of consistency with the continuous geodesic distances naturally arises.
While discrete geodesics are often inconsistent, "sampling theorems" have been derived guaranteeing the convergence of the discrete solution to the continuous one under certain conditions \cite{bernstein2000graph}. When a manifold is discretized as a graph satisfying these conditions, the discrete geodesic distances computed on it are guaranteed to converge to the continuous ones.
As an alternative to solving a discrete geodesic problem, another line of approaches suggests to discretize directly the partial differential equation governing the distance map on the underlying continuous domain,
$$\|\nabla \bb{d}\| = 1 \quad s.t. \quad \bb{d}|_{\tt source} = 0,$$
known as the eikonal equation.
Sethian~\cite{sethian1996fast} and Tsitsiklis~\cite{tsitsiklis1995efficient} have independently developed an efficient eikonal solver having  $O(n \log n)$ runtime, which is essentially a continuous variant of Dijkstra's algorithm.
This method, known as the \emph{fast marching method} (FMM), was later extended from regular grids to other types of graphs like triangular meshes \cite{kimmel1998fast} and point clouds \cite{memoli2005distance}, just to mention a few.
A faster version was introduced by Yatziv et al.\cite{yatziv2006n}, solving at $O(n)$ at the cost of quantizing the distances.

Even though the mentioned methods differ in the types of data handled and the specific calculations performed, they are all dynamic programming algorithms and share the property of gradually propagating the distance calculation from the source(s) to the rest of the vertices.
One potential advantage of this is that one can stop the calculation before covering the entire graph once the destination is achieved, saving time for `close' distances.
However, as a result of their inherent sequential structure and data-dependent access to memory, these methods are hard to parallelize and implement efficiently on modern computer architectures.
Furthermore, they usually cannot reuse the calculations from previous runs.

%-------------------------------------------------------------------------
%-------------------------------------------------------------------------
%-------------------------------------------------------------------------
\subsection{Stack based methods}

The other category of algorithms, in contrast to the former, operate on the entire vertex set or fixed subsets thereof.

One line of methods that provide a fast approximation of FMM are called \emph{fast sweeping}, and can be dated back to Danielsson \cite{danielsson1980euclidean} who proposed to calculate a complete Euclidean distance map using only four `sweeps' of a 2D plane in $O(n)$.
The method was later expanded to other types of manifolds, as formally suggested by Zhao~\cite{zhao2005fast}, after it was implicitly used in other works \cite{zhao2000implicit,tsai2002rapid,tsai2003fast}.
Even though these methods have $O(n)$ complexity, their main drawback is that for some manifolds it might require several repetitions to achieve sufficient accuracy,
up to $O(2^n)$ for an exact solution, as shown by Hysing et al.~\cite{hysing2005eikonal}.
A highly parallel variant called \emph{parallel marching} was suggested by Weber et al.~\cite{weber2008parallel}, who proposed to decompose the surface into regular grids, solving each in a manner similar to \cite{danielsson1980euclidean}. Even though parallel marching requires a low-distortion parameterization into quadrilateral patches, it is still one of the fastest implementations of surface distance computation to date.

The newest line of works in this category follows the heat method by Crane et al.~\cite{Crane}, which will be covered in greater detail in the sequel.

Finally, we would like to mention a recent work by Aflalo et al.~\cite{aflalo2013spectralMDS}, where spectral-based distance approximations are also performed as the input of a subspace parameterized multidimensional scaling (MDS) problem.
While this approach performs some approximation in a manner seemingly similar to ours, we will show in the sequel that it does not scale to general pairwise distance approximations.

%-------------------------------------------------------------------------
%-------------------------------------------------------------------------
%-------------------------------------------------------------------------

\section{Background}

%Before we introduce the heat method in the next section, we first cover some preliminaries of graph Laplacian.
Prior to detailing how the heat method works, we briefly review some preliminaries, mainly regarding graph Laplacian.

%-------------------------------------------------------------------------
%-------------------------------------------------------------------------
%-------------------------------------------------------------------------
%\subsection{Graph Laplacian and spectral analysis}

\paragraph{Graph Laplacian.}
\label{sec:laplacian}

%\subsubsection{Graph Laplacian}
Let $\mathcal{G}=(V,E)$ be an undirected graph with $|V|=n$ vertices. Let the graphs edges be further equipped with nonnegative weights $\{w_{ij}\}$, defining the $n\times n$ \emph{weighted adjacency matrix} of the graph as
$$%\begin{equation}
[\bb{W}]_{ij} =
\begin{cases}
w_{ij}, & \text{if } (i,j) \in E \\
0, & \text{otherwise}.
\end{cases}
$$%\end{equation}
Additionally, the \emph{vertex degree matrix} is an $n\times n$ diagonal matrix, whose elements are usually defined as
\begin{eqnarray}\label{eq:deg_mat}
	[\bb{A}]_{ii} = a_{ii} = \sum_j w_{ij}.	
\end{eqnarray}
%
%We define a \emph{location} on the graph as the minimal set of vertices needed to calculate gradient (locally). For example, two vertices on an edge, or 3 vertices on a face of in a triangular mesh.
%
The \emph{unnormalized} Laplacian of a graph is defined as the $n\times n$ matrix $\bb{L}_u = \bb{A} - \bb{W}$.
As with all Laplacians, it is easy to observe that the sum of all rows and columns is zero.
For an undirected graph, the Laplacian is a symmetric positive semidefinite matrix.
For other properties of the Laplacian, the reader is referred to \cite{mohar1991laplacian}.

One of the ways to normalize the Laplacian adopted here is the so-called \emph{random walk} Laplacian \cite{meila2001random,chung1997spectral}, defined as
\begin{equation}\label{eq:rw_lap}
\bb{L}_{rw} = \bb{A}^{-1}\bb{L}_u = \bb{I}- \bb{A}^{-1}\bb{W}.
\end{equation}
The name stems from its relation to the row-stochastic transition probability matrix
$\bb{P} =\bb{I}- \bb{L}_{rw}.$

\paragraph{Mesh Laplacian.}
A triangular mesh is a specific type of graph, realizing a homogeneous simplicial complex of order $2$.
Since meshes typically represent continuous geometric objects (manifolds), specific constructions of mesh Laplacians exist that consistently discretize some properties of the Laplace-Beltrami operator (LBO) of the underlying continuous manifold.
Mesh Laplacians have been extensively used in geometry processing; for a review, the reader is referred to \cite{sorkine2006differential}.

\begin{wrapfigure}[4]{C}{0.07\textwidth}
	\vskip-20pt
	\hskip-15pt
	\begin{overpic}[scale=1]{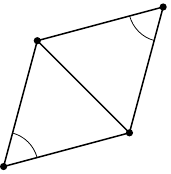}
		\put(15,20){$\alpha$}
		\put(60,65){$\beta$}
		\put(70,6){$i$}
		\put(8,70){$j$}
	\end{overpic}
\end{wrapfigure}
We adopt the popular \emph{cotangent weight scheme} first formalized in its current form by Meyer et al. \cite{meyer2003discrete}.
Under this scheme, the $n \times n$ \emph{stiffness matrix} $\bb{L}_c$ is constructed as an unnormalized Laplacian whose edge weights are
\begin{equation*}
w_{ij} = \frac{1}{2}\cot(\alpha) + \cot(\beta),
\end{equation*}
with $\alpha$ and $\beta$ being the two angles facing the edge $(i,j)$,
and the diagonal of degree matrix is the same as in \eqref{eq:deg_mat}.
This scheme also defines the \emph{mass matrix} $\bb{M}$ whose elements are $[\bb{M}]_{ii}=m_i$, where $m_i$ is $\tfrac{1}{3}$ the area of all triangles incident on vertex~$i$.
Finally, the mesh Laplacian is defined as $\bb{L}_m = \bb{M}^{-1} \bb{L}_c$.

\paragraph{Graph geodesics.}

A \emph{path} on the graph is a sequence of $p$ vertices $P=\{v_1,\ldots,v_p\} \in V$, where each consecutive pair is one of the graph edges, i.e. $(v_i,v_{i+1}) \in E$. %\: \forall \{i\}_1^{p-1}
The length of a path P is the sum of all its edge weights,
$$
L(P) = \sum_{i=1}^{p-1} w_{v_i,v_{i+1}}.
$$
A shortest path (also known as minimum geodesic on a continuous manifold) between two vertices $v$ and $v'$ is a path starting at $v_1 = v$ and ending at $v_p = v'$ whose length is minimum over all such paths.
%The \emph{geodesic} between a given pair of vertices is the path with the smallest length;
Note that a minimizer might not be unique as there might be several such paths.

As mentioned previously, sometimes graphs are used as a discretization of a continuous manifold.
In these cases the above definition of a geodesic has to be relaxed in order to allow convergence of the discrete solution to the continues one \cite{bernstein2000graph}.
Paths on a mesh, for example, will not be restricted to the edges and are allowed to traverse the faces.

%Let a \emph{simplex} on the graph be any set of vertices the can be use to measure intrinsic distance. For example, on a mesh these will be the faces, while in any general graph we will be limited to use the edges.
%Let a \emph{path} on the graph be any sequence of $p$ such simplices $\{s_1,\ldots,s_p\}$, where each consecutive pair is sharing at least one vertex.
%%, i.e. $(v_i,v_{i+1]}) \in E,\: \forall \{i\}_1^{p-1}$.
%The length of a path is the sum of all its edge weights.
%The \emph{geodesic} between a given pair of vertices with the smallest path length; note that the minimizer might not be unique as there might be several such paths.

\paragraph{Discretized gradient.}
We follow the definition of the intrinsic gradient for triangular meshes from \cite{Crane}, for the ease of comparison.
Other realizations exist, e.g., by Pokrass et al. \cite{pokrass2011correspondence}.

Let $f$ be a piece-wise linear function on the mesh defined by the values $f_i$ on the vertices. Let $\delta$ be a triangle formed w.l.o.g. by the vertices $1$, $2$ and $3$ with the corresponding function values $f_1, f_2$ and $f_3$. The gradient of $f$ on $\delta$ is constant and is defined by
%Let the triangle $s$ have its vertices $\{v_1,v_2,v_3\}$ be equipped with scalar values $\{f_1,f_2,f_3\}$ respectively. The gradient on $s$ is defined as
\begin{equation}
	\nabla \delta = \frac{1}{2\cdot\text{area}(\delta)}\sum_{i=1}^3 f_i \left({n}_s \times {e}_i\right)
\end{equation}
where ${n}_s$ is the normal to $\delta$, and ${e}_i$ is the edge opposing vertex $i$ oriented counterclockwise.

In a general (non-mesh) graph, we can no longer rely on faces, and define the gradient on edges. Specifically, an edge $e = (v_i,v_j)$ whose vertices have scalar values $\{f_i,f_j\}$ will have a gradient
$\nabla e_{ij} = \left( f_i-f_j \right) / w_{ij}$.

%-------------------------------------------------------------------------
%-------------------------------------------------------------------------
%-------------------------------------------------------------------------
\subsection{The heat method}
A recent seminal work by Crane et al. \cite{Crane} proposed a new approach to distance calculation dubbed as the \emph{heat method}.
This method, which is extended in this paper, is summarized in Algorithm \ref{alg:heat}.
While we focus on the specific realization of the heat method on triangular meshes, it is more general and can be applied to other types of data, as mentioned in \cite{Crane}.

\begin{algorithm}[t]
\SetKwInOut{Input}{input}
\SetKwInOut{Output}{output}

\hrulefill\\
\Input{Laplacian $\bb{L}$, Gradient $\nabla$, Divergence $\Delta$,\\ time value $t_0$ and origin point(s)}
\Output{Distance map $\bb{d}$}
\hrulefill

\parbox{.92\columnwidth}{
\begin{enumerate}[leftmargin=15pt]
\item Solve the heat equation \eqref{eq:heat} from the source point(s) for time value $t$, obtaining $\bb{u}_t$.
\item Calculate the gradient of the result $\bb{u}_t$, and normalize to unit length according to \eqref{eq:heat-nrm}, obtaining $\widehat{\bb{g}}$.
\item Solve Poisson equation \eqref{eq:poisson} for the divergence of the normalized gradient $\widehat{\bb{g}}$, obtaining $\bb{d}$.
\end{enumerate}}\\
\hrulefill\vspace{5pt}
 \caption{The Heat method \cite{Crane}}
 \label{alg:heat}
\end{algorithm}

The key idea behind the heat method stems from Varadhan's equation~\cite{varadhan1967behavior}:
it can be shown that the solution $u_t$ of the heat equation
\begin{equation}\label{eq:heat}
\left( \bb{M} + t \bb{L}_c \right) \bb{u}_t = \bb{u}_0
\end{equation}
describing the heat distribution at time $t$ is a function whose gradient direction coincides with that of a distance map from the same source (yet, unlike the distance map, the gradient magnitude $\| \nabla \bb{u}_t \|$ is not necessarily unit). The initial heat placement $\bb{u}_0$ at $t=0$ is a discretized Dirac distribution, having $1$ in the source location(s) and $0$ elsewhere.

Following this observation, we can normalize the gradient of $\bb{u}_t$ to unit length
\begin{equation}\label{eq:heat-nrm}
\widehat{\bb{g}} = \frac{\nabla \bb{u}_t}{\|\nabla \bb{u}_t \|},
\end{equation}
and solve Poisson's equation for the distance map $\bb{d}$ from the same source such that its Laplacian coincides with the divergence of the normalized gradient $\widehat{\bb{g}}$
\begin{equation}\label{eq:poisson}
\bb{L}_c \bb{d}= {\nabla}\cdot\widehat{\bb{g}}.
\end{equation}
%where d denotes the distance map from the same source.

We did not specify here the exact implementation of the divergence operator for triangular mesh, and refer the reader to Section 3.2 in \cite{Crane} for details.
Note however, that the discretization or even a consistent definition of this operator on more general graphs is a limitation of the heat method.
In what follows, we describe an alternative technique for distance map calculation similar in spirit to the heat method, yet relying only on the gradient but not on the divergence.

%-------------------------------------------------------------------------
%-------------------------------------------------------------------------
%-------------------------------------------------------------------------
\section{Proposed method - SpectroMeter}
The proposed method which we dub \emph{SpectroMeter} can be thought of as a variant of the heat method \cite{Crane} in the spectral (frequency) domain. The details of the method are given in the following section and are summarized in Algorithm \ref{alg:spectral}.
The proposed method relies on several spectral properties of the Laplacian that we briefly highlight next.

%As mentioned, this method follows the lines of the heat method \cite{Crane}, doing so in the spectral domain. This section lays out the details of the suggested method, which are also summarized in Algorithm \ref{alg:spectral}.

\begin{algorithm}[t]
	\SetKwInOut{Input}{input}
	\SetKwInOut{Output}{output}
	
	\hrulefill\\
	\Input{Laplacian eigenfunctions $\bb{\Phi}$, their gradient $\nabla\bb{\Phi}$, $O(k)$ locations, origin point(s)}
	\Output{Approximate distance map}
	\hrulefill\\
	\parbox{.92\columnwidth}{
		\begin{enumerate}[leftmargin=15pt]
			\item Calculate the heat kernel $\bb{h}_t$ from the source points at the given $O(k)$ locations on the mesh, according to \eqref{eq:spectral_heat}.
			\item Calculate $\nabla\bb{h}_t$ the gradient of $\bb{h}_t$ at these locations and normalize to unit length $\widehat{\bb{g}} = \nabla \bb{h}/{\|\nabla \bb{h}\|}$ .
			\item Find coefficients $\bb{a}$ that best approximate $\widehat{\bb{g}}$, i.e. $\widehat{\bb{g}} \approx \sum_{i=1}^k a_i \nabla\bb{\phi}_i$, according to \eqref{eq:l2_coeffs}.
			\item Set the first coefficient(s) $a_i$ corresponding to constant $\bb{\phi}_i$'s so that the result is non-negative (or has zero value at the source).
			\item Synthesize approximate distance at desired location(s) $x$ according to $d(x) = \bb{\phi}(x)\bb{a}$.
		\end{enumerate}
	}\\
	\hrulefill\vspace{5pt}
	\caption{SpectroMeter - the proposed sublinear distance map approximation algorithm. The full (non-sublinear) flavor of this algorithm simply uses \emph{all} the graph points rather than the given $O(k)$ samples.}
	\label{alg:spectral}
\end{algorithm}

\paragraph{Laplacian Spectrum.}
%The proposed method relies on several spectral properties of the Laplacian that we briefly highlight next.
Spectral graph theory is a vast field with numerous and ramified applications \cite{mohar1991laplacian,cvetkovic1980spectra}.
%Spectral analysis of the Laplacian is not a new topic \cite{mohar1991laplacian,cvetkovic1980spectra}, and has many applications in the literature.
For example, a recent tutorial by von Luxburg~\cite{von2007tutorial} covers applications of different types of Laplacians to clustering and graph-cuts.
Being essentially an averaging operator, graph Laplacian excels at promoting smooth functions on the graph \cite{sorkine2006differential}, and specifically, it has been shown that the Laplacian eigenbasis is optimal in the $\ell_2$ sense for the approximation of smooth functions~\cite{aflalo2015optimality}.

We denote the eigendecomposition of a Laplacian $\bb{L}$ as $\bb{L}\bb{\phi}_i = \lambda_i \bb{\phi}_i$, with $\lambda_i$ and $\bb{\phi}_i$ being the eigenvalues and the corresponding eigenfunctions.
We will assume ordering by increasing value of $\lambda_i$.
It is easy to show that the smallest eigenvalue $\lambda_0$ has the value $0$ with a constant corresponding eigenvector and has multiplicity equal to the number of connected components in the graph.

%-------------------------------------------------------------------------
%-------------------------------------------------------------------------
%-------------------------------------------------------------------------
\paragraph{Preprocessing.}
As with the other distance methods \cite{sethian1996fast,Crane}, some of the calculations have to be done for any input, and can be pre-computed and stored to reduce the amortized cost of distance computation.

The proposed technique, being a spectral method, relies on first performing eigendecomposition of the graph Laplacian.
Specifically, we require the first smallest $k \ll n$ eigenvalues and the corresponding eigenfunctions. These low-frequency harmonics can be reused for many other tasks relying on the Laplacian spectrum. We henceforth denote the truncated eigenbasis collectively by the matrix $\bb{\Phi} = (\bb{\phi}_1, \dots, \bb{\phi}_k)$.
Next, we also calculate the (intrisic) gradients of all the eigenvectors (the constant one(s) have vanishing gradients), denoting the results collectively as $\nabla \Phi = (\nabla \phi_1,\dots,\nabla \phi_k)$.

Finally, we sample the graph at $O(k)$ locations, preferably using farthest point sampling \cite{eldar1997farthest} (FPS) based on an `exact' method like FMM.
This sampling is used for the sublinear version of the algorithm, as detailed in the sequel. % (covered in Section \ref{sec:sublinear}).
%One can look at this step as similar to S-MDS by Aflalo et al....\cite{aflalo2013spectralMDS}

%-------------------------------------------------------------------------
%-------------------------------------------------------------------------
%-------------------------------------------------------------------------
\paragraph{Heat in the spectrum.}
The major difference between \cite{Crane} and the proposed method is that the solution is done in the spectral domain. This has two implications.
First, the heat equation \eqref{eq:heat} can be solved using the eigenbasis.
One way to do this is to apply the Woodbury matrix inversion identity \cite{woodbury1950inverting} to \eqref{eq:heat}, which will still cost at least $O(n^2)$ and might be inaccurate.
We chose to approximate the solution directly through the spectrum (following, e.g., \cite{sun2009concise}) as the truncated series
\begin{equation}\label{eq:spectral_heat}
\hat{h}_t(x_s,y) = \sum_{i=1}^k e^{-\lambda_i t} \bb{\phi}_i(x_s) \bb{\phi}_i(y).
\end{equation}
The function $\hat{h}_t(x_s,y)$ approximates as the \emph{heat kernel}, which measures the amount of heat that flows from a source point $x_s$ to point $y$ after time $t$.
For the case of multiple (or non-point) sources, the solution for a set of source points $\chi_s=\{x_i\}_{i=1}^s$ is the average of the heat kernel of all source points, $\hat{h}_t(\chi_s,y) = \tfrac{1}{s}\sum_i \hat{h}_t(x_i,y)$.
We will henceforth denote the values of $\hat{h}_t(x_s,y)$ at all vertices $y$, and averaged over all sources $\chi_s$ in the graph as $\bb{h}_t$.

In essence \eqref{eq:spectral_heat} is the spectral approximation of the solution of the `exact' heat diffusion \eqref{eq:heat}, but is more efficient as it can be done in $O(nk)$ instead of $O(n^2)$ with $k \ll n$. The approximation is accurate as long as the time constant $t$ is sufficiently big and the residual of the truncated series \eqref{eq:spectral_heat} is small.
This formulation allows further acceleration down to $O(k^2)$ as described in the sequel.

The second implication is that we no longer need to solve the full $n \times n$ Poisson equation \eqref{eq:poisson}, but rather find a set of $k$ Fourier coefficients $\bb{a}$ defining a low-rank approximation of the function $\hat{\bb{g}} = \bb{\Phi} \bb{a}$, such that its gradient $\nabla \hat{\bb{d}} = \nabla\bb{\Phi} \bb{a}$ is the closest in the $\ell_2$ sense to the normalized gradient $\hat{\bb{g}} = \nabla \bb{h}_t/\| \nabla \bb{h}_t \|$ of the approximate heat kernel, resulting in the following closed-form expression
\begin{equation}\label{eq:l2_coeffs}
\bb{a}^* = \underset{\bb{a}}{\mathrm{argmin}} \|\widehat{\bb{g}} - \nabla\bb{\Phi}\bb{a}\|_2 = (\nabla\bb{\Phi})^\dagger\widehat{\bb{g}},
\end{equation}
where $^\dagger$ marks pseudo inverse.
Note that this formulation does not rely on the divergence operator.
Since the coefficient(s) of the constant eigenfunction(s) cannot be determined from \eqref{eq:l2_coeffs} (as the corresponding gradients vanish), we set them to make the entire map non-negative or zero at the source.

\paragraph{Random walk.} The only difference for general (non-mesh) graphs is that the heat equation is replaced with random walk process, where the notion of `heat' is replaced with probability of arrival at a location $y$ after $t$ steps. We follow the definition from \cite{von2007tutorial} for this probability
\begin{equation}\label{eq:spectral_rw}
\widehat{h}_t(x_s,y) = \sum_{i=1}^k \left(1- \lambda_i \right)^t  \bb{\phi}_i(x_s) \bb{\phi}_i(y),
\end{equation}
where here $\{\lambda_i,\bb{\phi}_i\}$ are the eigenvalues and the corresponding eigenfunctions of the random-walk laplacian $\bb{L}_{rw}$.

%-------------------------------------------------------------------------
%-------------------------------------------------------------------------
%-------------------------------------------------------------------------
\paragraph{Sublinear approximation.}
\label{sec:sublinear}
Both the approximation of the heat kernel \eqref{eq:spectral_heat} and the solution of \eqref{eq:l2_coeffs} are $O(nk)$.
However, we observe that both \eqref{eq:spectral_heat} and \eqref{eq:l2_coeffs} are highly over-determined, in the case of $k\ll n$, and still can be solved unambiguously if only a subset of $O(k)$ graph vertices is considered. This reduces the overall complexity of the method to $O(k^2)$, which is sublinear in $n$. This excludes, of course, the pre-computation mentioned above which costs no less than $o(nk)$ for eigen decomposition, then  $O(nk\log n)$ for FPS, and finally $O(k^3)$ for the matrix inversion in \eqref{eq:l2_coeffs}.

It is important to emphasize that this sublinear complexity is for a single pairwise distance, and increasing the number of sources or destinations will increase the overall complexity.
For example, approximating the distance from one point to the entire vertex set will still cost $O(nk)$.
That said, empirical results show that our method is still considerably faster than all mentioned methods, even in the latter case.

%The question now is what locations to choose, and how.
%If this was done in the Euclidean domain, we would be trying to recover a band unlimited function from a
%Without going into details

In practice we sample the graph using FPS based on one of the `exact' method mentioned previously.
The minimum amount of samples is the number of non-constant eigenfunctions.
We use an amount of distinct samples slightly higher than what makes $\nabla \bb{\Phi}$ in \eqref{eq:l2_coeffs} numerically full rank.

A similar sampling is performed by Aflalo et al.~\cite{aflalo2013spectralMDS}, but is not limited by the basis size $O(k)$, nor can it be generalized to full distance computation,
%as will be shown in the experimental section below.
as shown below.

%-------------------------------------------------------------------------
%-------------------------------------------------------------------------
%-------------------------------------------------------------------------
\paragraph{Finding optimal value of $t$.}
In \cite{Crane}, the time value $t$ in the heat equation was set to be the squared average edge length, based on discretization arguments.
This selection seem to work in practice for most shapes even though there is no worst-case guarantee, and still failing in some isolated cases (e.g. in Figure~\ref{fig:teaser} we had to use ten times the latter value).

\begin{figure}
	\centering
	\includegraphics[width=.9\linewidth]{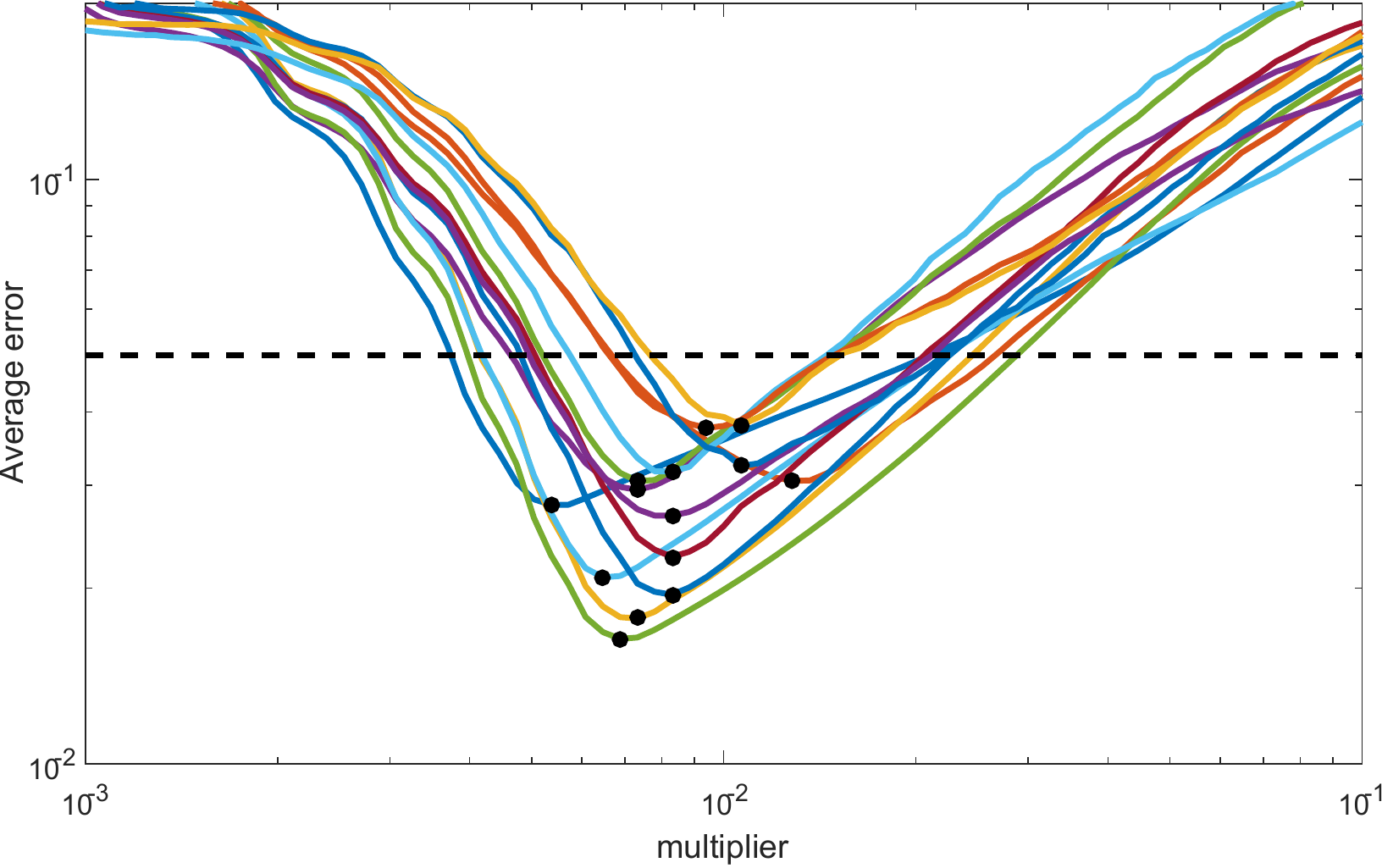}
	\caption{Average error (measured relative to FMM and normalized by the shape diameter) as a function of the multiplier $m$ in the selection of the time parameter $t = m \, \tr(\bb{M})$.
		Each curve corresponds to a shape set from Table I in \cite{Crane} or from the TOSCA dataset~\cite{bronstein2008numerical}.
		Notice that in most examples a uniform setting $m=8\times 10^{-3}$ is close to the shape-specific optimal parameter value (black dots) and yields mean error below $5\%$ (dashed line). This figure was inspired by its counterpart in \cite{Crane}. }
	\label{fig:time_mutl}
\end{figure}

As shown in \cite{bronstein2008numerical}, the spectral heat kernel in \eqref{eq:spectral_heat} does not depend on the discretization, but does not scale trivially with the global scaling of the shape.
%This makes the selection of the best $t$ not as straightforward, and we leave a principled selection of its value to future work.
A scale agnostic version of \eqref{eq:spectral_heat} would then be
\begin{equation}\label{eq:spectral_heat_area}
\hat{h}_t(x_s,y) = \tr(\bb{M})\sum_{i=1}^k e^{-\lambda_i t \tr(\bb{M})} \bb{\phi}_i(x_s) \bb{\phi}_i(y),
\end{equation}
where $\tr(\bb{M})$ is the shape area.
In a manner similar to \cite{Crane}, we also choose define the time value $t$ by empirically setting a multiplier $m$
\begin{equation}\label{eq:time_mult}
t = m \, \tr(\bb{M}).
\end{equation}
Testing on several shapes from various sources, we set $m=8\times 10^{-3}$, as can be drived from Figure~\ref{fig:time_mutl}.
Note, that while in \cite{Crane} the time value is discretization dependent, our time value depends only on the total area of the shape.

%We propose an exhaustive but fast search to estimate this time value.
%Since the previously mentioned $O(k)$ locations are a result of FPS, one can also obtain an exact $k\times k$ pairwise distance matrix between those locations.
%We can now run our method several times on these $k$ points, keeping the value of $t$ producing distances most similar to the FPS run.
%In practice, sufficient accuracy was achieved by about $15$ log-spaced samples in the range between $10$ and $10^4$ times the average edge length. %\textbf{TEST FOR SHAPE AREA}

 \paragraph{Comparison to spectral MDS.}
 \begin{wrapfigure}[10]{C}{0.15\textwidth}
 	\vskip-13pt
 	\hskip-15pt
 	\begin{overpic}[scale=.3,natwidth=389,natheight=456]{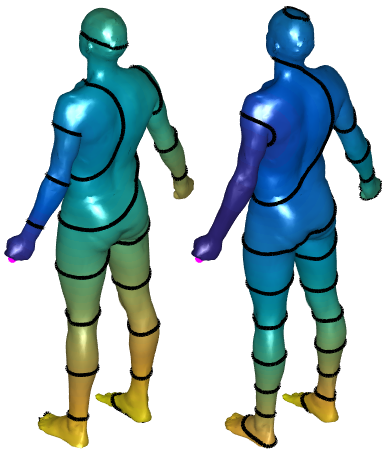}
 		\put(53,-8){\small SMDS \cite{aflalo2013spectralMDS}}
 		\put(07,-8){\small FMM \cite{kimmel1998fast}}
 	\end{overpic}
 \end{wrapfigure}
 As mentioned previously, one part of the Spectral MDS (SMDS) method \cite{aflalo2013spectralMDS} includes distance calculations seemingly similar the SpectroMeter.
 This part of the SMDS algorithm includes interpolation of distance maps sampled on a sparse set of points, based on a projection on the Laplacian eigenbasis. While \cite{aflalo2013spectralMDS} report very low reconstruction errors, these are true only for maps whose origin is close to one of the former samples. If one would use the same method with origin far from these samples, the result will look similar to the result of \eqref{eq:spectral_heat}, i.e. non-unit magnitude gradients with the correct direction (as can be seen in the inset).%Figure~\ref{fig:SMDS}.

 %\begin{figure}
 %	\centering
 %	\begin{overpic}[scale=.5]{SMDS.png}
 %		\put(60,-5){SMDS \cite{aflalo2013spectralMDS}}
 %		\put(15,-5){FMM}
 %	\end{overpic}
 %	\vspace{5pt}
 %	\caption{Two distance map on a shape from FAUST~\cite{FAUST2014}, whose origin is on the left hand.
 %		Shown are FMM and the distance interpolation form \cite{aflalo2013spectralMDS}. The origin is not one of the sample points used by \cite{aflalo2013spectralMDS}, resulting in a map similar to heat kernel \eqref{eq:spectral_heat}.
 %	}
 %	\label{fig:SMDS}
 %\end{figure}

\begin{table}
	\centering   \addtolength{\tabcolsep}{-1pt}
	\begin{tabular}{lcccccc}
		\hline
		basis size & 25 & 50 & 100 & 250 & 350 & 500 \\
		\hline
		\hline
		relative & 4.28 & 2.51 & 1.79 & 0.77 & 0.60 & 0.43 \\
		$\ell_2$ & 1.36 & 0.87 & 0.64 & 0.29 & 0.24 & 0.17 \\
		$\ell_\infty$ & 4.90 & 3.18 & 2.39 & 1.39 & 1.15 & 0.96 \\
		\hline%\vspace{-1pt}
	\end{tabular}
	\caption{Average error of projecting FMM distance on the Laplacian basis of the cat shape ($26 \times 10^3$ vertices) from TOSCA \cite{bronstein2008numerical} shown in Figure~\ref{fig:teaser}. Shown for different number of harmonics are the $\ell_2$ and  $\ell_\infty$ errors in \% of intrinsic diameter as well as the relative error in \%. The $\ell_\infty$ error does not decay fast due to the Gibbs phenomenon~\cite{hewitt1979gibbs}. The errors are averaged over several runs with difference source points. %\textbf{TODO - add a plot showing residual error on the shape}.
	}
	\label{table:recon_errs_proj}
\end{table}

%-------------------------------------------------------------------------
%-------------------------------------------------------------------------
%-------------------------------------------------------------------------
\section{Experimental evaluation}

In what follows, we assess the actual performance of the discussed approximation method compared to the `exact' methods, and highlight some of its limitations. Two flavors of SpectroMeter will be tested: \emph{full}, which uses the entire vertex set, and \emph{sublinear}, which uses only the predefined $O(k)$ samples.
%\subsection{Data}
%FAUST , TOSCA, kids? ,
%MNIST? , CIFAR?
For comparison, we used a Matlab mex implementation of FMM from the authors of \cite{kimmel1998fast}, and a Matlab implementation of the Heat method accelerated by the CHOLMOD algorithm \cite{chen2008algorithm}, used originally in \cite{Crane}.
We chose not to compare to the exact geodesic distance on polyhedra \cite{mitchell1987discrete} in the large scale experiments, as its results are very similar to FMM result (less than $0.5\%$ error), while the computational time is prohibitively slow.

Errors are measured using the $\ell_2$ and $\ell_\infty$ norms, normalized by the graph diameter.
Another criterion we measure is relative error, defined as $\tfrac{|d-d_{ref}|}{d_{ref}+\epsilon}$ for a result $d$ versus a reference value of $d_{ref}$.
We emphasize that latter error is somewhat biased \emph{against} our method: errors of spectral representation usually concentrate in points where the function is non-differentiable, e.g., at the sources, where the relative error is most sensitive (small denominator). Nevertheless, we include these statistics here as it was used in previous studies  \cite{aflalo2013spectralMDS,Crane,surazhsky2005fast} to which we compare.

%-------------------------------------------------------------------------
%-------------------------------------------------------------------------
%-------------------------------------------------------------------------
\subsection{Spectral representation accuracy}
First, we test the accuracy of projecting (essentially, compressing) distance maps on the Laplacian eigenbasis.
Table~\ref{table:recon_errs_proj} shows reconstruction error of FMM distance maps, averaged over the shapes from Figure~\ref{fig:time_mutl}, using different numbers of harmonics. As suggested by \cite{aflalo2015optimality}, the $\ell_2$ error is decaying very fast, and is quite negligible even for $k=100$ harmonics.  We attribute this to the fast decay of coefficients of distance maps, which can be shown to be quadratic for Euclidean domains \cite{AbsT}.
The $\ell_\infty$ error does not decay as fast, which can be related to the \emph{Gibbs phenomenon}~\cite{hewitt1979gibbs}.

Next, for comparison, we repeat the experiment only this time using SpectroMeter sublienar (full has very similar stats).
The result, summarized in Table \ref{table:recon_errs_ours}, suggest that our method achieves maximum performance around $k=250$ harmonics after which diminishing returns are observed. This as well, may be due to the fast decay of coefficients \cite{AbsT}.
Since this result was consistent across many shapes we tested, in all the following experiments we set $k=250$.

\begin{table}
	\centering   \addtolength{\tabcolsep}{-1pt}
	\begin{tabular}{lcccccc}
		\hline
		basis size & 25 & 50 & 100 & 250 & 350 & 500 \\
		\hline
		\hline
		relative & 55.86 & 32.41 & 15.80 & 9.93 & 9.74 & 9.81 \\
		$\ell_2$ & 19.33 & 13.63 & 6.35 & 2.84 & 2.88 & 3.10 \\
		$\ell_\infty$ & 46.17 & 37.05 & 22.14 & 7.46 & 6.88 & 7.65 \\
		\hline%\vspace{-1pt}
	\end{tabular}
	\caption{Average error of our method compared to FMM, in the same setting as Table~\ref{table:recon_errs_proj}. It can be seen that our method achieves its best performance at $k=250$, and does not improve with more harmonics, which can be attributed to the fast decay of coefficients \cite{AbsT} in the presence of noise. The relative error is high since spectral reconstruction has high error around non-differentiable areas like the source.}
	\label{table:recon_errs_ours}
\end{table}

%-------------------------------------------------------------------------
%-------------------------------------------------------------------------
%-------------------------------------------------------------------------
\subsection{Complexity analysis}
\begin{figure}
\centering
\includegraphics[width=.9\linewidth]{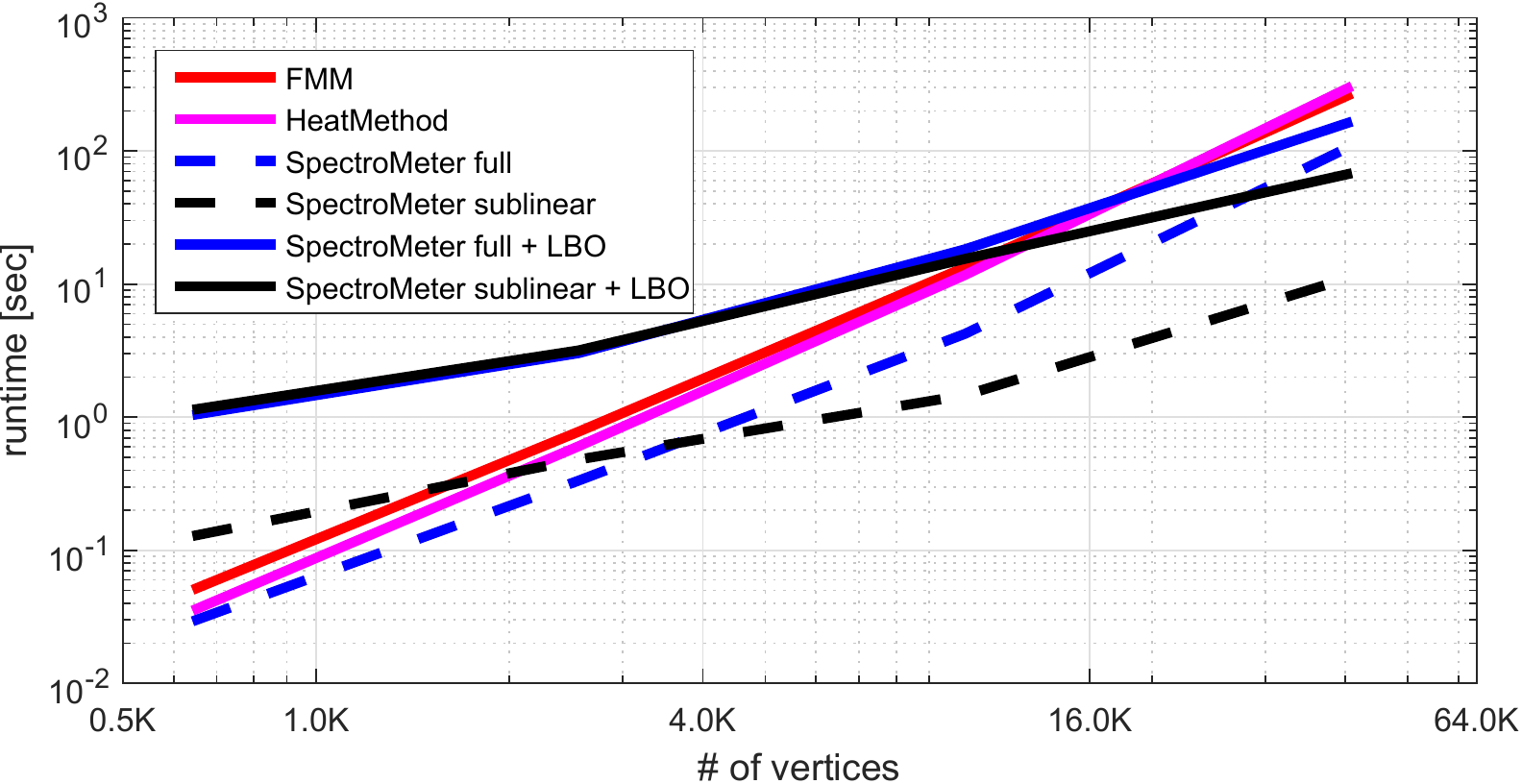}
\caption{
	Runtime of different methods, on different graph sizes.
	Presented is the time for calculating pairwise distance from 15\% of the vertices to all the rest on a sphere mesh.
	Shown are: FMM, Heat method and SpectroMeter, both full and sublinear.
	Even though this task has quadratic complexity, our method scales better than the exact ones, even when counting the complexity of basis construction (marked `LBO' in the plot).}
\label{fig:time}
\end{figure}

In order to measure how all methods scale with the graph size, we ran them on an implicit shape discretized as a mesh, with gradually increasing vertex count. For each mesh, we measured the distance from $5\%$ of the vertices to all the rest. Note that due to the size of the result, this task has quadratic complexity in the size of graph, even for our sub-linear method.
As can be seen in Figure~\ref{fig:time}, our approach outperforms the other ones, and the incline of the sub-linear flavor of our method is the lowest. If we include the eigendecomposition into the timing, both SpectroMeter implementations are still faster for large graphs.

Table~\ref{table:errs_complexity} shows error of the latter experiment compared to FMM. While the errors of SpectroMeter are roughly twice as high as the heat method, it scales more gracefully for larger graphs.
%SHOW peak memory usage? errors?

\begin{table}
	\centering
	\addtolength{\tabcolsep}{-1pt}
	\begin{tabular}{llcccc}
		\hline
		&\# vertices & 0.6K & 2.5K & 10K & 20K \\
		method &err type & & & & \\
		\hline
		\hline
		heat& relative & 3.24 & 2.19 & 1.52 & 1.11 \\
		method & $\ell_2$ &1.35 & 0.92 & 0.66 & 0.51 \\
		\cite{Crane} & $\ell_\infty$ &1.95 & 1.42 & 1.08 & 0.86 \\
		\hline
		& relative & 6.25 & 6.53 & 6.32 & 6.09 \\
		full & $\ell_2$ &2.37 & 2.39 & 2.27 & 2.16 \\
		& $\ell_\infty$ &2.92 & 3.41 & 3.69 & 3.82 \\
		\hline
		& relative & 6.23 & 6.53 & 7.17 & 6.77 \\
		sublinear & $\ell_2$ & 2.36 & 2.39 & 2.26 & 2.13 \\
		& $\ell_\infty$ & 2.92 & 3.40 & 3.67 & 3.79 \\
		\hline
	\end{tabular}
	\caption{Errors of distance methods w.r.t. to an FMM result on the same corpus, measured in $\%$ of the intrinsic diameter of a sphere mesh with different vertex count.}
	\label{table:errs_complexity}
\end{table}

%-------------------------------------------------------------------------
%-------------------------------------------------------------------------
%-------------------------------------------------------------------------
\subsection{Distance on a nearest neighbor graph}

One of the contributions of this work is that we do not need do define a divergence operator.
This allows us to apply the method on non-manifold graphs, for example a graph generated by running $k$ nearest neighbors ($k$-NN) on data from a Euclidean domain.
We generated a graph from $7$-NN of $3\times 10^3$ points in $\Rr^5$, which resulted in a graph Laplacian with similar sparsity to the meshes in other experiments.

\begin{figure}
	\centering
	\addtolength{\tabcolsep}{-2pt}
	\begin{tabular}{cccc}	
		\vspace{5pt}
		\includegraphics[width=.2\linewidth,natwidth=747,natheight=717]{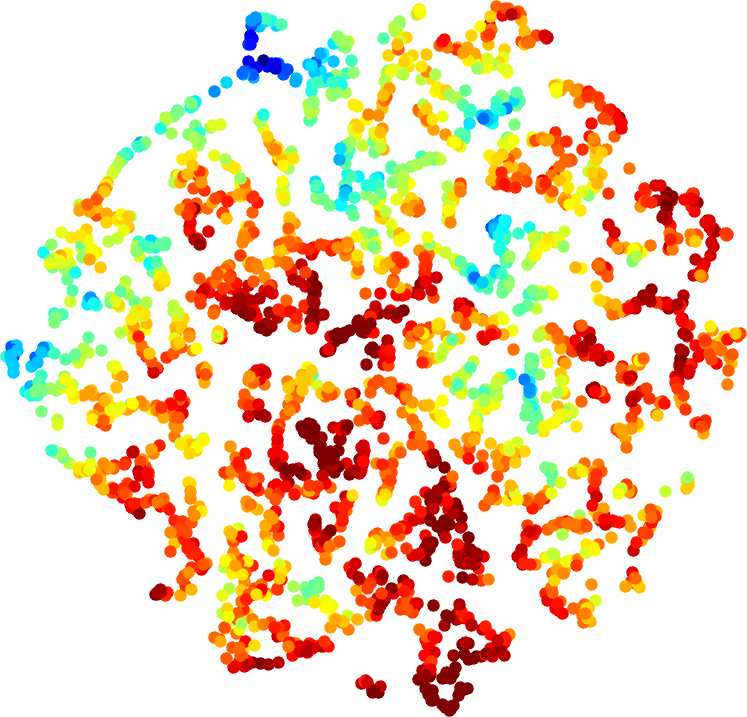} &
		\includegraphics[width=.2\linewidth,natwidth=747,natheight=717]{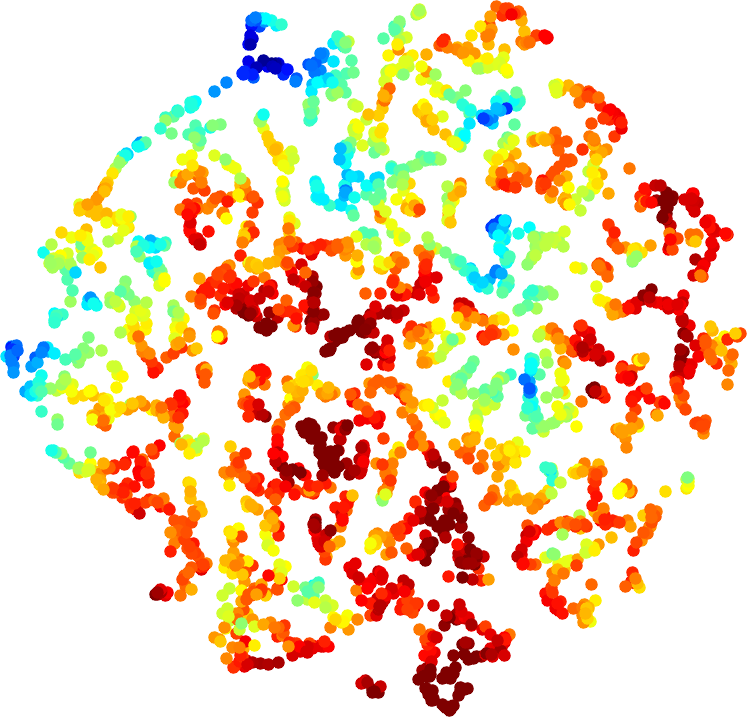} &
		\includegraphics[width=.2\linewidth,natwidth=747,natheight=717]{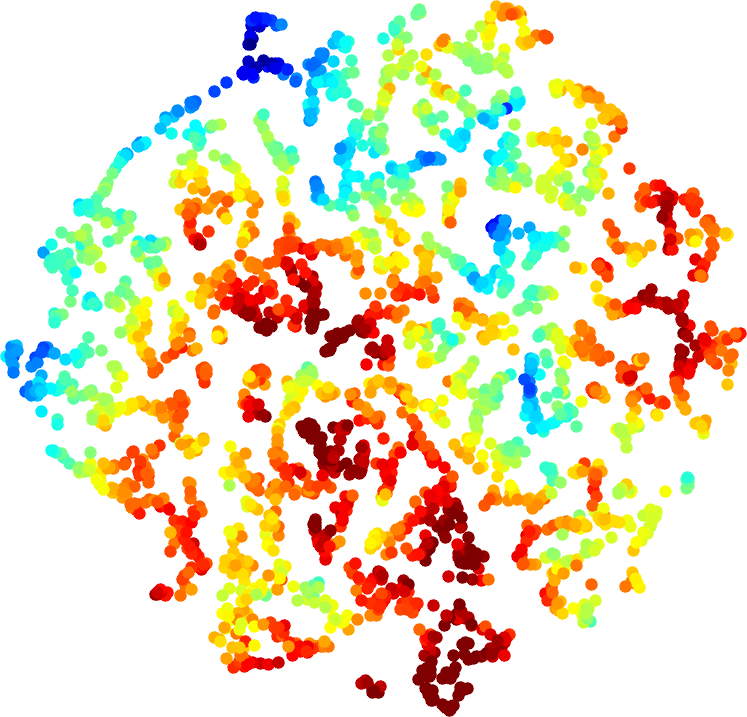} &
		\includegraphics[width=.2\linewidth,natwidth=747,natheight=717]{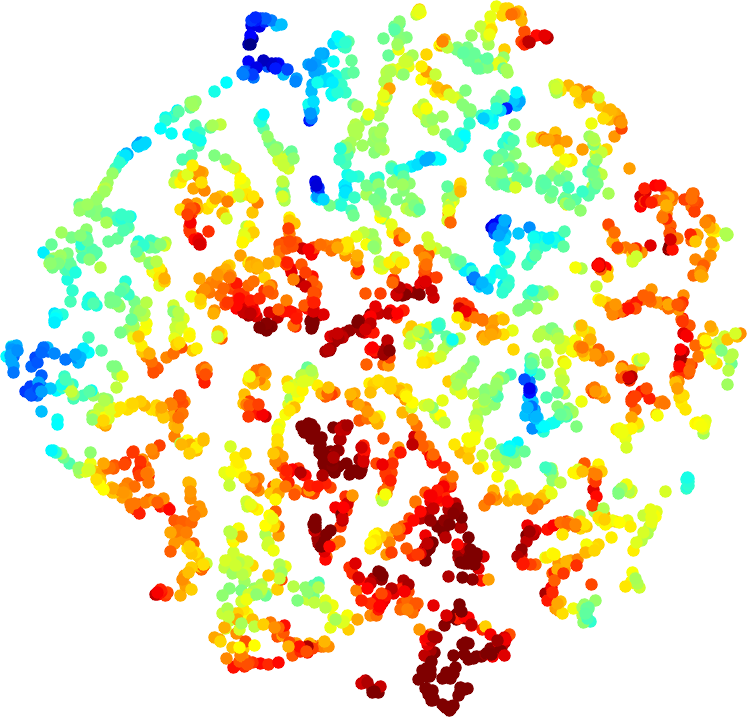} \\
		\vspace{5pt}
		\includegraphics[width=.2\linewidth,natwidth=747,natheight=717]{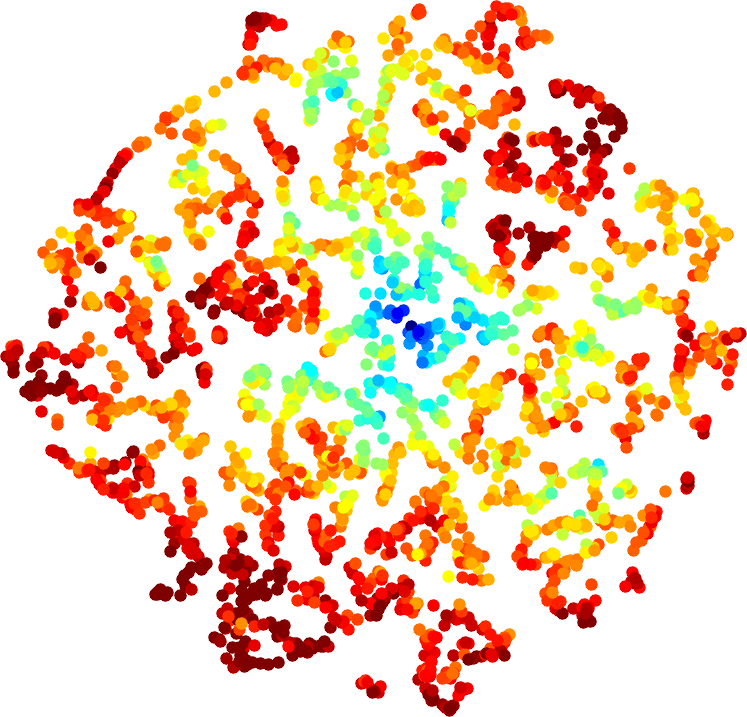} &
		\includegraphics[width=.2\linewidth,natwidth=747,natheight=717]{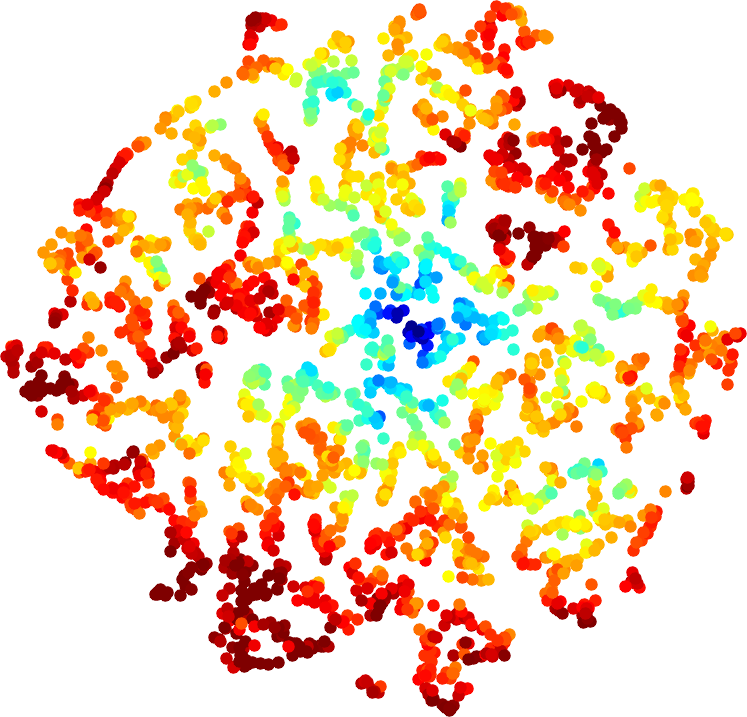} &
		\includegraphics[width=.2\linewidth,natwidth=747,natheight=717]{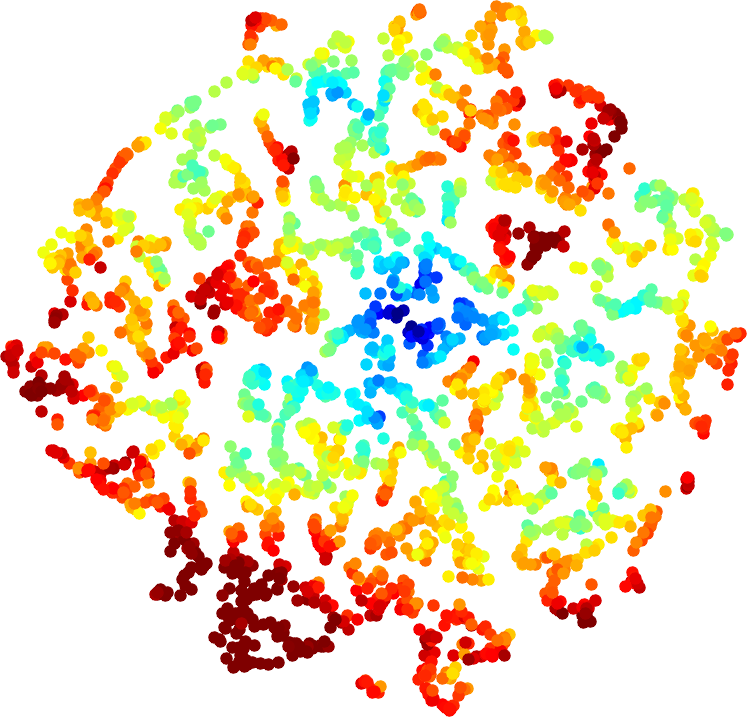} &
		\includegraphics[width=.2\linewidth,natwidth=747,natheight=717]{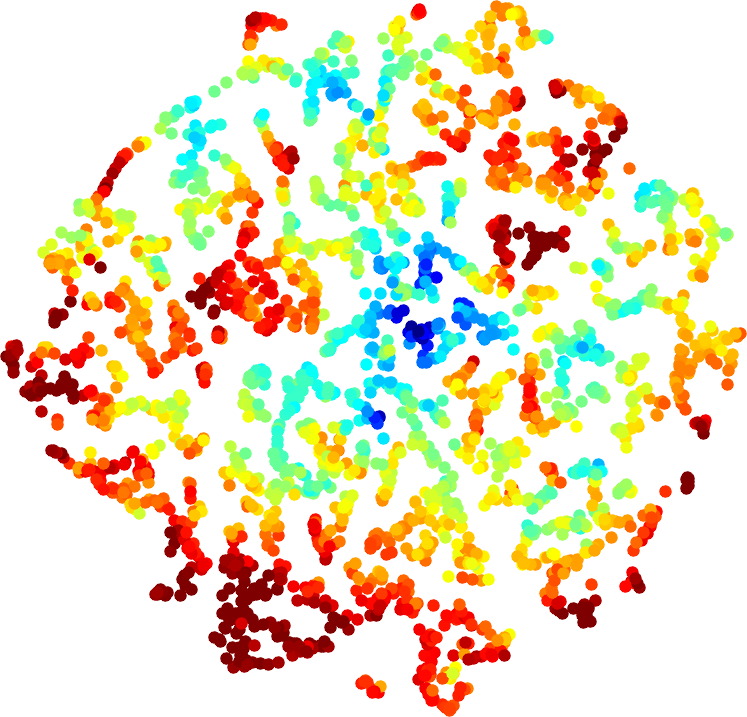} \\
		\vspace{5pt}
		\includegraphics[width=.2\linewidth,natwidth=747,natheight=717]{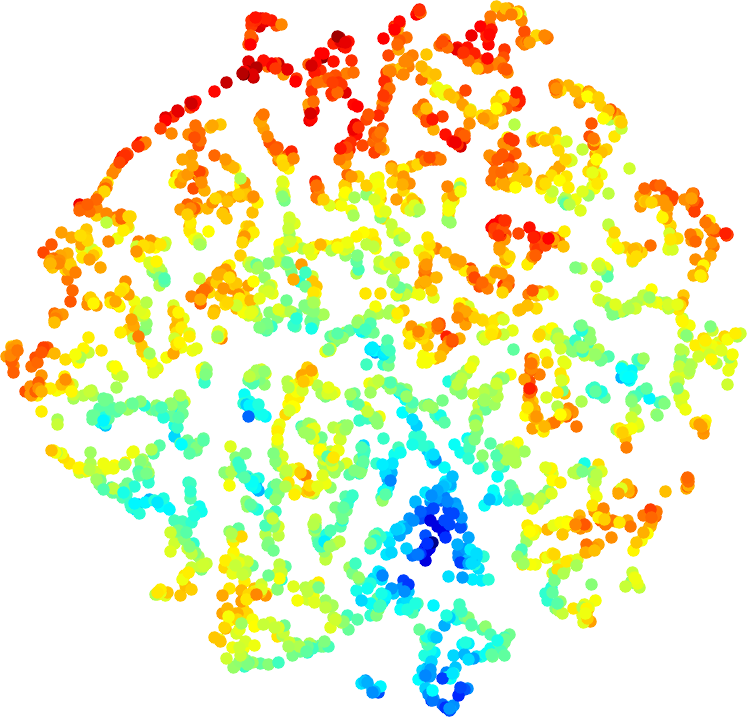} &
		\includegraphics[width=.2\linewidth,natwidth=747,natheight=717]{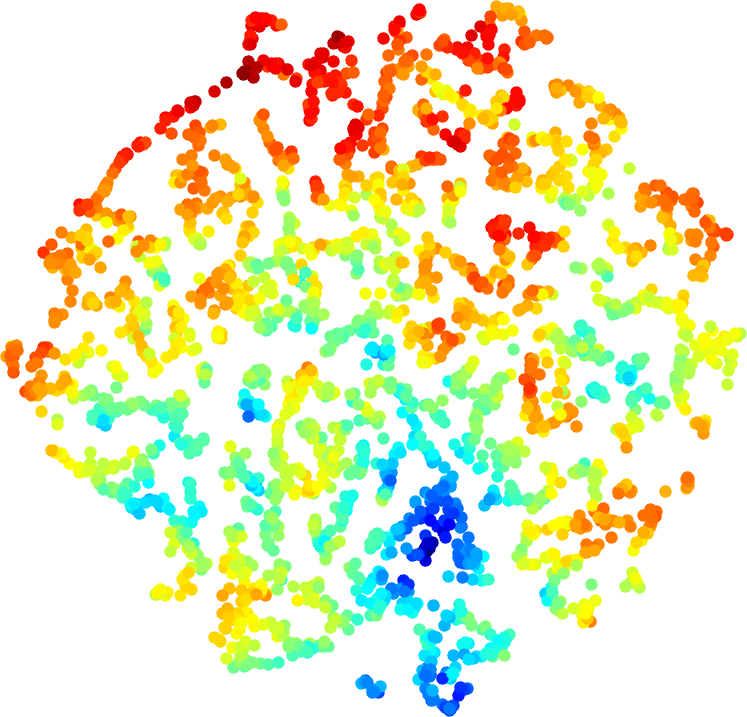} &
		\includegraphics[width=.2\linewidth,natwidth=747,natheight=717]{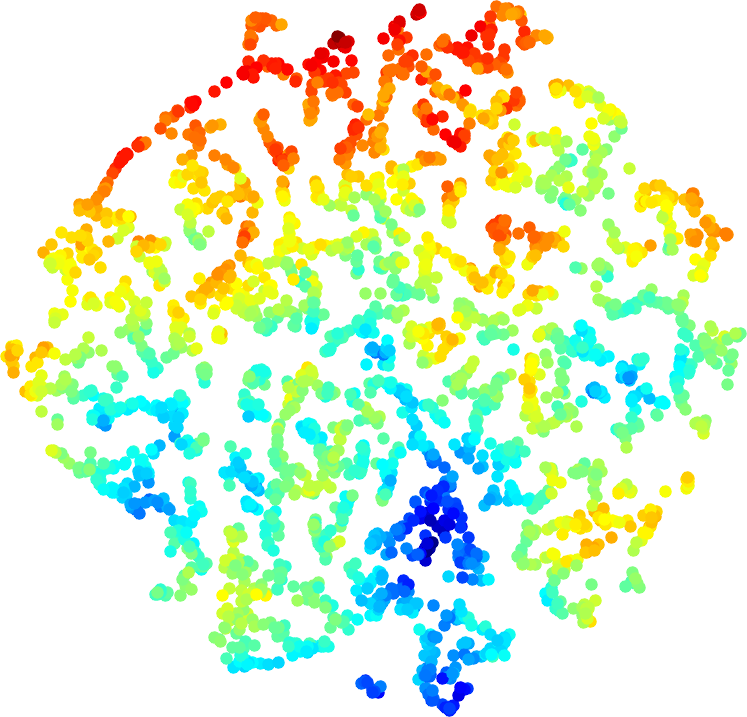} &
		\includegraphics[width=.2\linewidth,natwidth=747,natheight=717]{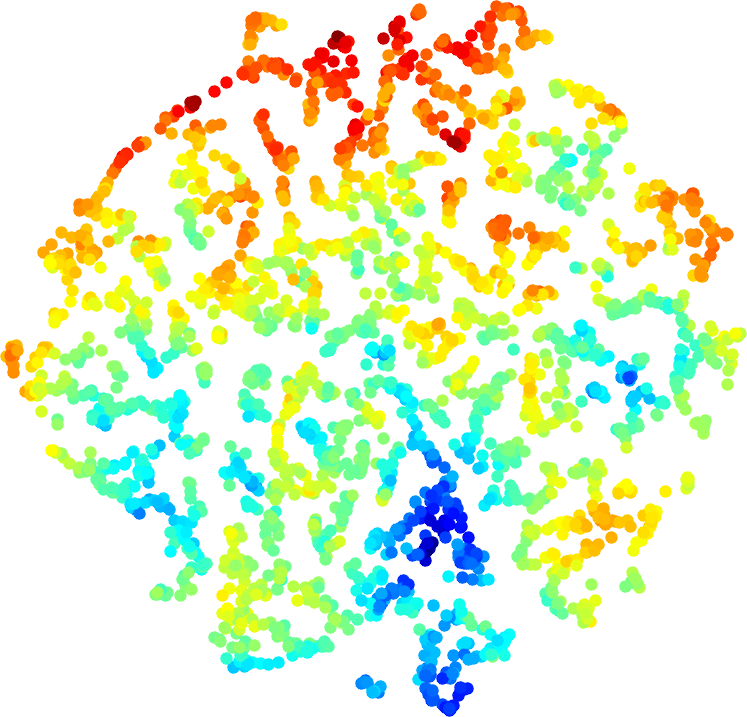} \\	
		%\small
		& Spectrally-   &  &\\
		Exact & projected  & SpectroMeter  & SpectroMeter\\
		Dijkstra \cite{dijkstra1959note} & Dijkstra \cite{dijkstra1959note}  & Full & Sublinear\\
	\end{tabular}
	\caption{Distance maps produced by four methods on a graph created using $7$-NN in $\Rr^{5}$.
		The graph vertices are visualized in 2D using t-SNE \cite{maaten2008visualizing}.
		Shown in each row is a results for a different source, color mapped from blue (small distance) to red (large distance). }
	\label{fig:knn}
\end{figure}

Figure \ref{fig:knn} depicts distance maps from three source points, visualized in two dimensions using t-SNE \cite{maaten2008visualizing}.
The resulting maps are very similar to the exact ones generated using Dijkstra's method \cite{dijkstra1959note}, but require a fraction of the computational cost.

In order to give a qualitative measure for the results, we sorted the vertices according to the resulting distance in each map separately, and measured Kendel's $\tau$ \cite{kendall1938new} as the permutation distance
\begin{equation}
\tau = \frac{\text{\# discordant pairs}}{n\left(n-1\right)/\:2},
\end{equation}
compared to Dijkstra's result.
On average, projection onto the Laplacian eigenbasis gave $12\%$ error, while both types of SpectroMeter gave $14\%$ error.
While this may seem high, note that random guess results in $50\%$ error, and the probability of randomly achieving error below $30\%$ vanishes quickly even for permutations with $n=50$.
%TODO - add Heat-like-method here as well?

% MNIST?
% CIFAR?(based on GIST)
% maybe show quick-shift?

%-------------------------------------------------------------------------
%-------------------------------------------------------------------------
%-------------------------------------------------------------------------
\subsection{Shape correspondence}
We now show our methods' result on a recent method for shape correspondence by Vestner et al.~\cite{vestner2016bayesian}. Due to lack of space, this section is not self-contained, as we only cover the parts relevant for this work. We refer the reader to the aforementioned manuscript for details. In a nutshell (skipping all theoretic justification), the latter method finds bijective correspondence by solving a linear assignment problem (LAP) over all the vertices
\begin{equation}
\bb{\Pi}^* = \min_{\bb{\Pi}\in {\mathcal P}} \tr\left(\exp(-\bb{D}_2 / \sigma^2)\bb{\Pi}_0\bb{D}_1\bb{\Pi}\right) ,
\end{equation}
where $\bb{D}_1$ and $\bb{D}_2$ are pairwise distance matrices, $\exp$ is element-wise exponent, $\bb{\Pi}_0$ is an initial correspondence guess (possibly non bijective) and ${\mathcal P}$ is the set of all possible permutations of size $n$.
LAP has an $O(n^3)$ worse-case complexity, but can be solved in $O(n^2 \log n)$ operation on average using the auction method \cite{auction} (taking, e.g., a few seconds for a problem of size $n=10^3$).
In practice, the smooth geometry of the problem allows further multi-scale acceleration, which makes the LAP cost matrix sparse (zero entries indicate forbidden matches). In the particular case of triangular meshes, truncating the pairwise distance matrices at $5\%$ of the diameter we can achieve about $99\%$ sparsity, allowing solution of a large LAP problem of size $n=10^4$ in about one minute.
The main bottleneck that remains is the calculation of the full pairwise distance matrices.

We tested the correspondence method on the FAUST dataset~\cite{FAUST2014} containing 3D scans of human shapes with $7\times10^3$ vertices.
We solve LAP using the sparse auction solver publicly provided by the authors of \cite{auction_fast}.
The initial guess input  $\bb{\Pi}_0$ was a obtained by using the current state-of-the-art correspondence method by Rodol\`{a} et al.~\cite{rodola2015point} on a coarse version of the shape with $2\times10^3$ vertices, and interpolating the result to full resolution using nearest neighbors.
Note that the initial guess method \cite{rodola2015point} also uses the Laplacian eigendecomposition, making Spectrometer's complexity even more attractive in this case:
Average run times in seconds (excluding LAP) is $260$ for FMM~\cite{kimmel1998fast}, $120$ for the heat method~\cite{Crane}, compared to $42$ and $24$ respectively for the full and sublinear flavors of SpectroMeter.
%260.68  121.1  42.744  26.513

\begin{figure}
	\centering
	\includegraphics[width=.9\linewidth]{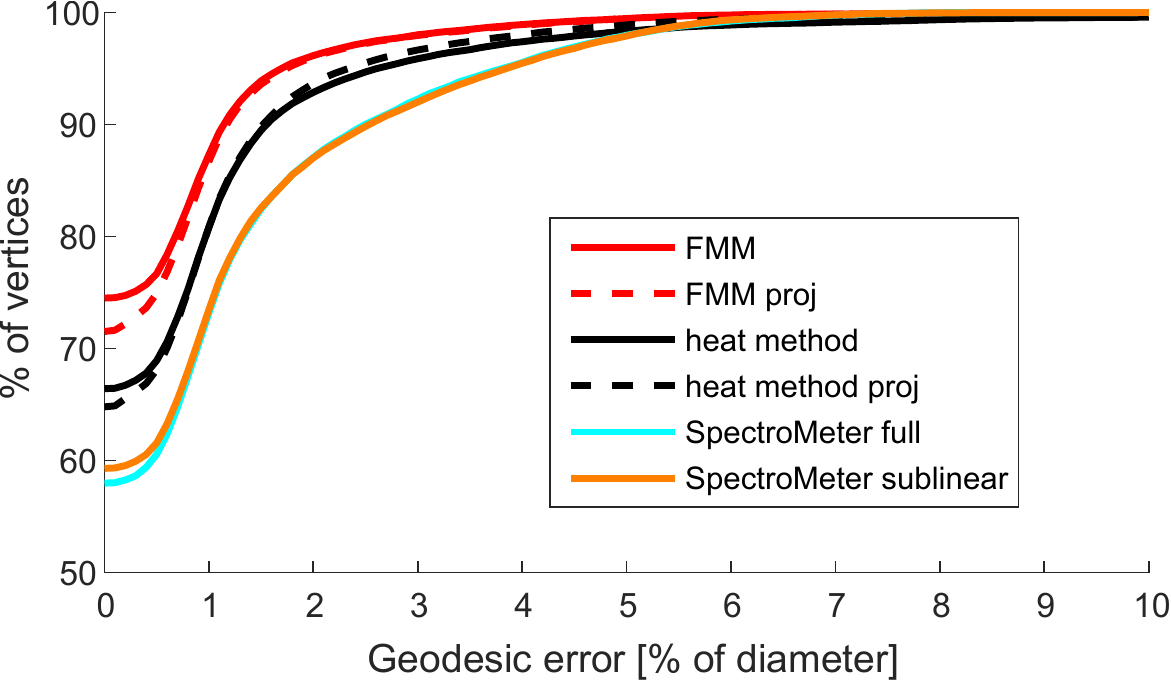}
	\caption{Performance of the shape correspondence method from \cite{vestner2016bayesian} on the FAUST dataset~\cite{FAUST2014},
		based on different distance methods as input. Dashed lines indicate that the distance was projected onto the LBO basis (i.e. compressed).
		SperctroMeter yields slightly lower results due to the fact it is an approximation, but can further scale up to very big shapes.}
	\label{fig:corr}
\end{figure}

Figure \ref{fig:corr} shows results of the method \cite{vestner2016bayesian} using several distance methods as input.
We also tested the effect of projecting the distance matrix onto the Laplacian eigenbasis.
While the performance is lower for Spectrometer compared to the exact distance, the overall method of \cite{vestner2016bayesian} becomes more scalable as it can be run for shapes of very high vertex count.

%-------------------------------------------------------------------------
%add tri-mesh experiminets on S-MDS?
%Multi-scale functional maps?
%-------------------------------------------------------------------------

%\subsection{Missing experiments}
%Failure cases of Aflalo, Crane, Ours.
%Show Gibbs on the shape, perhaps on a line / cross-section.

%-------------------------------------------------------------------------
\section{Conclusion}
We presented an algorithm allowing to approximate pairwise distances on sparsely-connected graphs (such as, for example, triangular meshes) in sub-linear time given the low-frequency eigenfunctions of the Laplacian. Empirical evidence exhibits very good $\ell_2$ approximation errors and reasonable $\ell_\infty$ errors.

%\paragraph{limitations}
Similarly to both the heat method \cite{Crane} and FMM \cite{sethian1996fast}, the obtained distance map might violate the triangle inequality or distance symmetry.
For similar reasons, it is possible that these distance maps cannot be used to compute the geodesics themselves via back-tracking, due to possible existence of local extrema along the path.

%\paragraph{future work}
The influence of boundary conditions that has been ignored here deserves attention.
%Additionally, we would like to devise a principled way or even a closed-form expression for the time value in the heat kernel \eqref{eq:spectral_heat}.
Additionally, we would like to devise a criterion more robust than $\ell_2$ for the solution of \eqref{eq:l2_coeffs}.
Lastly, but most importantly, rigorous theoretical bounds on the average and worst-case approximation errors shall be developed.

%-------------------------------------------------------------------------
\subsection*{Acknowledgments}
RL is supported by the European Google PhD fellowship in machine learning. AB is supported by ERC StG RAPID 335491. The authors are grateful to Dana Berman, Or Litani, and the anonymous reviewers, for \emph{many} helpful pointers.

%%the appendix is before references due to the 8 page limit (excluding references)
%\input{6_appendix.tex}

{%\small

%	\bibliographystyle{./3dv_stuff/ieee}
%	\bibliography{SpectroMeter}
}

\end{document}